\documentclass[aps,floats,twocolumn]{revtex4}
\usepackage{epsf}
\usepackage{amsfonts, bm}
\usepackage{amsmath}
\usepackage{amssymb}
\usepackage[latin2]{inputenc}
\usepackage[T1]{fontenc}
\usepackage{epsfig}
\usepackage{mathrsfs}
\usepackage{verbatim}

\newcommand{\tr}{\text{tr}}

\newcommand{\h}{\mathcal{H}}
\newcommand{\de}{\text{d}}
\newcommand{\Po}{\mathcal{P}_1}

\newlength{\dinwidth}                         
\newlength{\dinmargin}
\DeclareMathAlphabet{\scr}{U}{rsfs}{m}{n} 

\begin{document}

\title{Group-theoretical approach to entanglement}

\author{J. K. Korbicz$^{1,2}$ and M. Lewenstein$^{2,1}$}

\affiliation{$^1$ Institut f\"ur Theoretische Physik, Universit\"at Hannover, Appelstr. 2, D-30167
Hannover, Germany}

\affiliation{$^2$ ICREA and ICFO--Institut de Ci\`{e}ncies Fot\`{o}niques, Mediterranean Technology Park, 08860
Castelldefels (Barcelona), Spain}

\begin{abstract}
We present a novel, universal description of quantum entanglement using group theory and non-commutative characteristic functions. It leads to new reformulations of the separability problem as well as allows us to generalize it, thus connecting theory of entanglement and harmonic analysis. As an example, we translate and analyze the positivity of partial transpose (PPT) criterion and a simple criterion for pure states into the group-theoretical language. We also show that when applied to finite groups, our formalism embeds separability problem in a given dimension into a higher dimensional but highly symmetric one. Finally, our formalism reveals a connection between the very existence of entanglement and group non-commutativity.
\end{abstract}

\maketitle

\section{Introduction}
Despite numerous attempts \cite{PPT} -- \cite{Otfredo} and substantial time passed after its formulation \cite{Werner}, the problem of efficient description of entangled states of multipartite systems  --- the, so called, separability problem --- remains still open. We recall \cite{Werner, primer, Nielsen} that a state $\varrho$ of an $N$-partite system is called separable if it can be represented as a convex combination of product states:
\begin{equation}\label{Nsep}
\varrho=\sum_i p_i |\psi^{(1)}_i\rangle\langle\psi^{(1)}_i|\otimes\dots\otimes |\psi^{(N)}_i\rangle\langle\psi^{(N)}_i|.
\end{equation}
Otherwise $\varrho$ is called entangled. The importance of the separability problem lies in both practical applications, connected to the quantum information processing \cite{Nielsen},  as well as in the conceptual issues of quantum mechanics (see e.g. Refs. \cite{concept, CBH}).

In this work we present a novel, group-theoretical approach to the separability problem in finite dimensions. Our approach is based on a generalization of standard characteristic functions and results from the following two observations: i) it is possible to identify a bipartite Hilbert space $\mathbb{C}^m\otimes\mathbb{C}^n$ with a tensor product of representation spaces of two independent irreducible unitary representations of some suitable compact group $G$; ii) one can then perform a non-commutative Fourier transform \cite{Folland} and assign to each density matrix a unique function on $G \times G$, satisfying certain positivity conditions. This function is an analog of a classical characteristic function \cite{AbrSteg}, but is defined on a generically non-Abelian group. We will call such functions ``non-commutative characteristic functions''. The group $G$ will be called the ``kinematical group'' of an individual system. 

Let us emphasize that for a generic quantum system the choice of the kinematical group is at this stage arbitrary --- the {\it only} requirement is that $G$ should possess irreducible unitary representations, matching the dimensionality of the system's Hilbert space. This freedom makes our approach very flexible and allows us to use as $G$ e.g. finite groups as well as Lie groups.

The potential significance of non-commutative characteristic functions for quantum mechanics has been first, up to our knowledge, pointed out in the physical literature by Gu in Ref. \cite{Gu}. However, no investigation of the separability problem has been carried out there, as the work of Gu predates the seminal paper of Werner \cite{Werner}. On the other hand, standard characteristic functions played a crucial role in solving the separability problem for Gaussian states \cite{Giedke}, and in studying quantumness of states of harmonic oscillator (see Ref. \cite{chuj} and references therein).

In the present work we use non-commutative characteristic functions to restate the separability problem in a new language. Although we do not present any new entanglement tests, our results offer a new point of view on this long-standing problem, and link it to harmonic analysis and group theory. In particular, we pose a generalized separability problem for non-commutative characteristic functions, which is an interesting mathematical problem in itself. As an example of the necessary generalized separability criterion, we reformulate the PPT criterion \cite{PPT} in group-theoretical terms and show that it is connected to a certain simple operation on non-commutative characteristic functions. This connection is universal, and holds irrespectively of the group used. Apart from that, we translate one of the necessary and sufficient criteria for pure states. Using the freedom in the choice of the kinematical group, we examine an interesting case of finite kinematical groups, like permutation groups. This leads to an embedding of the separability problem in a given dimension into a higher dimensional one, with some specific symmetries, however. Quite interestingly, we also show a {\it purely formal} similarity of our formalism to local hidden variables (LHV) models \cite{WernerWolf}. Finally, the conceptually attracting feature of our approach is, that it allows to connect the very existence of entanglement with the group non-commutativity. 

The work is organized as follows: in Section II we define the non-commutative characteristic functions, and review their properties (using Ref. \cite{Folland} as the main mathematical reference). In Section III we use the developed formalism to reformulate the separability problem. Section IV is dedicated to the study of the PPT criterion. We show its ``robustness'' with respect to (w.r.t.) the change of the kinematical group. We also briefly examine one of the criteria for pure states there. In Section V we examine our separability criterion on finite groups. In Section \ref{lhv} we remark on the connection to the LHV models. Finally, in Section \ref{noncom} we sketch a possible reformulation of the mathematical language of quantum statistics, exposing the connection between entanglement and non-commutativity of the kinematical group.

\section{Non-commutative characteristic functions}
We begin with presenting the general set up of our work. We consider an arbitrary compact group $G$ (it may or may not be a Lie group) and let $\tau$ be any of its irreducible, unitary representation (in the sequel by representation we will always mean a unitary representation) acting in a Hilbert space $\mathcal{H}_\tau$. We study linear operators $A$ acting in $\mathcal{H}_\tau$, and, in particular, density matrices $\varrho$. In the present work we fix $G$ to be compact, as we study only finite dimensional systems here, and for a compact group all of its irreducible unitary representations are necessarily finite dimensional (see Refs. \cite{Gu, chujnia} for formalism of non-commutative characteristic functions on non-compact groups). Following Gu \cite{Gu} (see also Ref. \cite{Barnum}), we assign to each operator $A$ a continuous complex function $\phi_A$ on $G$ through:
\begin{equation}\label{phi}
\phi_A(g):=\tr\big[A \tau(g)\big]\,.
\end{equation}
For the particular case of a density matrix $\varrho$, the function $\phi_\varrho$ is a non-commutative analog of the usual Fourier transform of a probability measure -- if we think of a state $\varrho$ as of a quantum analog of a classical probability measure \cite{Mackey}, then $\phi_\varrho$ is an analog of its characteristic function. Indeed, from the positivity of $\varrho$ and Eq. (\ref{phi}), it follows that \cite{Gu}: 
\begin{equation}\label{fdt}
\iint_{G \times G} \de g \,\de h \overline{f(g)}\phi_\varrho(g^{-1}h)f(h)\ge 0 \ \text{for any}\ f\in L^1(G),
\end{equation}
where $\de g$ is a normalized Haar measure on $G$, and the bar denotes complex conjugation \cite{l1}. Functions $\phi:G \to \mathbb{C}$ satisfying the above property are called positive definite on $G$ \cite{Folland}. Moreover, $\phi_\varrho$ is normalized:
\begin{equation}\label{norm}
\phi_\varrho(e)=1
\end{equation}
($e$ denotes the neutral element of $G$), which follows from the normalization of $\varrho$. Thus, $\phi_\varrho$ possesses all the features of a classical characteristic function, but it is defined on a non-Abelian group. Hence the term ``non-commutative characteristic function''. It is the main object of our study. 

Note that characteristic functions (\ref{phi}) are generally easy to calculate explicitly. For example, when $G=SU(2)$ and $\tau=\tau_j$ carries spin $j$, they are polynomials of degree $2j$ in the group parameters. As an example, we calculate in Appendix A the characteristic function for the $3\otimes 3$ Horodecki's state from Ref. \cite{Pawel}.

The crucial point for our approach is that, since $\tau$ is irreducible, one can invert the non-commutative Fourier transform (\ref{phi}) and recover operator $A$ from its characteristic function \cite{Gu, Folland}:   
\begin{equation}\label{ro}
A=\int_G \de g\,\,d_\tau\, \phi_A(g) \tau(g)^\dagger ,\quad d_\tau:=\text{dim}\h_\tau.
\end{equation}
The proof of (\ref{ro}) is most easily obtained by taking the matrix elements of both sides in some orthonormal basis of $\h_\tau$, and then by using the orthogonality of the matrix elements of $\tau$, guaranteed by the Peter-Weyl Theorem \cite{ort}. An interesting implication of Eq. (\ref{ro}) is that multiplication of operators corresponds to taking convolutions of the corresponding functions (\ref{phi}):
\begin{equation}
\phi_{AB}=d_\tau \phi_A * \phi_B,\label{splot}
\end{equation}
where we define $f * f' (g):= \int_G \de h f(h)f'(gh^{-1}) = \int_G \de h f(h^{-1}g)f'(h)$ (in the last step we substituted $h\to h^{-1}g$ and used the fact that $\de g^{-1}=\de g$ for compact groups \cite{Folland}). In particular, state $\varrho$ is pure iff:
\begin{equation}
\phi_\varrho=d_\tau \phi_\varrho * \phi_\varrho.\label{czysta}
\end{equation}

Let us now focus on the space of all normalized, positive definite functions on $G$, i. e. the space of all continuous functions $\phi$, satisfying the conditions (\ref{fdt}) and (\ref{norm}). We denote this space by $\mathcal{P}_1(G)$. It is a convex subset of the space of all continuous functions on $G$, and the set of its extreme points we denote by $\mathcal{E}_1(G)$. The structure of $\mathcal{E}_1(G)$ is described by the Gelfand-Naimark-Segal (GNS) construction (see e.g. \cite{Folland} or the note \cite{GNS}): $\phi\in\mathcal{E}_1(G)$ iff there exists an irreducible unitary representation $\tau_\phi$ of $G$ and a normalized vector $\psi_\phi\in\h_{\phi}$ (the space of $\tau_\phi$) such that $\phi(g)=\langle \psi_\phi |\tau_\phi(g) \psi_\phi\rangle$. Thus, every $\phi\in\mathcal{E}_1(G)$ is a characteristic function of some pure state $\psi_\phi \in \mathcal{H}_{\phi}$. In particular, because of Eq. (\ref{czysta}), it satisfies: $\phi=d_{\tau_\phi} \phi * \phi$.

Obviously, $\mathcal{P}_1(G)$ contains more functions than just characteristic functions of the type (\ref{phi}). To identify which $\phi\in\mathcal{P}_1(G)$ are characteristic functions of states, first note that from Eq. (\ref{phi}) it follows that:
\begin{equation}\label{span} 
\phi_\varrho(g)=\sum_i p_i \langle \psi_i|\tau(g) \psi_i\rangle,
\end{equation}
where we used any convex decomposition (for example an eigenensemble) of $\varrho$: $\varrho=\sum_i p_i |\psi_i\rangle\langle \psi_i|$. From Eq. (\ref{span}), we see that the decomposition of $\phi_\varrho$ into extreme points from $\mathcal{E}_1(G)$ contains only one, fixed representation $\tau$. Conversely, let $\phi=\sum_i p_i \phi_i$ where $\mathcal{E}_1(G)\ni\phi_i = \langle \psi_i |\tau \psi_i\rangle$ for each $i$ (such sums are finite, since all irreducible representations are finite-dimensional), then $\phi=\phi_\varrho$, where $\varrho:= \sum_i p_i |\psi_i\rangle\langle \psi_i|$.  

Let us describe the inverse non-commutative Fourier transform (\ref{ro}) of the whole $\Po (G)$. Note that since $G$ is compact, the set of its irreducible representations is discrete, and we can label them by some natural index $k$. Then, any $\phi\in\Po (G)$ defines, through the integral (\ref{ro}), a positive semidefinite operator $\varrho_k(\phi)$ for every irreducible representation $\tau_k$. To prove it, note that for any $\psi\in\mathcal{H}_{k}$ (the space of $\tau_k$) holds:
{\setlength \arraycolsep{0pt}
\begin{eqnarray}
\langle \psi|&\varrho_k(\phi)& \psi\rangle = \int \de g\,\,d_k\, \phi(g) \langle \psi|\tau_k(g)^\dagger \psi\rangle \nonumber\\
&=& \iint \de h\,\, \de g\,\,d_k\, \phi(h^{-1}g) \langle \tau_k(h)^\dagger \psi|\tau_k(g)^\dagger \psi\rangle\nonumber\\
&=& \sum_{\mu=1}^{d_k} \iint \de h\,\,\de g\,\,d_k\, \overline{\langle e_\mu |\tau_k(h)^\dagger \psi\rangle}\,\phi(h^{-1}g)\nonumber\\
& &\times \langle e_\mu |\tau_k(g)^\dagger \psi\rangle\ge 0\label{p3}.
\end{eqnarray}}
In the second step above we changed the variables $g\to h^{-1}g$, used the invariance of $\de g$ and inserted $1=\int_G \de h$, since the integrand did not depend on $h$. Then we inserted a unit matrix, decomposed w.r.t. an arbitrary basis $\{e_\mu\}$ of $\mathcal{H}_{k}$. However, generically $\varrho_k(\phi)$ is subnormalized as for a generic $\phi$ there appear all irreducible representations of $G$ in the convex decomposition of $\phi$ into $\mathcal{E}_1(G)$. Hence, each  $\phi\in\Po (G)$ defines a positive semidefinite operator in the space $\bigoplus_k \mathcal{H}_{k}$, where the sum is over all irreducible representations of $G$, through:
\begin{equation}\label{ro1}
\varrho(\phi):=\bigoplus_k \varrho_k(\phi),   
\end{equation}
while each $\varrho_k(\phi)$ is given by Eq. (\ref{ro}). Only such defined operator $\varrho(\phi)$ is normalized, which follows from the identity $\phi(g)=\sum_k \tr\big[\varrho_k(\phi)\tau_k(g)\big]$ \cite{Folland}. Of course, if $\tau_k$ is not present in the decomposition of  $\phi$ then, from Peter-Weyl Theorem, $\varrho_k(\phi)=0$ \cite{ort}.

Summarizing, states on an irreducible representation space $\mathcal{H}_\tau$ are in the one-to-one correspondence with functions on $G$ satisfying the conditions (\ref{fdt}), (\ref{norm}), and (\ref{span}) \cite{GNS}. The last condition ensures that in the decomposition (\ref{ro1}) there appears only representation $\tau$, and hence the operator given by Eq. (\ref{ro}) acts in the desired space and is normalized. The correspondence $\varrho \leftrightarrow \phi_\varrho$ may be heuristically viewed as a change of basis: $|e_\mu\rangle\langle e_\nu| \leftrightarrow \langle e_\nu |\tau(\cdot) e_\mu \rangle$. 

The presented formalism is closely related to that of generalized coherent states \cite{Perelomov}. Within the latter, every density matrix on $\h_\tau$ can be represented as: $\varrho=\int_{G/H}\!\de x\, P_\varrho (x) |x\rangle\langle x|$, where $H$ is an isotropy subgroup in representation $\tau$ of some fixed vector $\psi_0\in\h_\tau$ and $|x\rangle$ are the corresponding coherent states. However, unlike non-commutative characteristic function, $P$-representation $P_\varrho$ is generally non-unique and do not encode positivity of a density matrix in a simple manner. For applications of generalized coherent states to the study of entanglement see e.g. Refs. \cite{Barnum}, \cite{Braunstein}, \cite{Karol}.

\section{Application to the study of entanglement}
Having established the formalism, we proceed to reformulate the separability problem in terms of non-commutative characteristic functions. Let us consider a bipartite, finite dimensional system, described by a Hilbert space $\mathcal{H}:=\mathbb{C}^m\otimes\mathbb{C}^n$. At this point, we arbitrarily identify the spaces $\mathbb{C}^m$ and $\mathbb{C}^n$ with two independent representation spaces $\h_\pi$, $\h_\tau$ of irreducible representations $\pi$, $\tau$ of some compact kinematical group $G$: 
\begin{equation}\label{ident}
\mathbb{C}^m \equiv \h_\pi, \quad \mathbb{C}^n \equiv \h_\tau.
\end{equation}
Of course, the group and the representations should be chosen to match the desired dimensions $m$, $n$. As we mentioned in the Introduction, this is the only constraint we impose on $G$. 

The identification (\ref{ident}), although mathematically always possible and non-unique, may seem arbitrary from the physical point of view. For instance, for a given system we could have chosen another kinematical group $G'$, possessing suitable representations. This freedom may in fact turn out to be a big advantage of the formalism, as the choice of $G$ can be optimized in each practical case. There is also a ``universal'' kinematical group $G=SU(2)$ --- since it possesses irreducible representations in all possible finite dimensions, it can serve as a kinematical group for all finite dimensional systems. The results of the previous Section imply then that we can describe through the formulas (\ref{phi}) and (\ref{ro}) {\it all} states in {\it all} finite dimensions in terms of non-commutative characteristic functions on $SU(2)$. Thus, without a loss of generality, we may always treat our system as a system of (possibly artificial) independent spins $j_1:=(m-1)/2$ and $j_2:=(n-1)/2$.

Having done the identification (\ref{ident}), we can view the Hilbert space of the  full system $\mathcal{H}=\mathcal{H}_\pi\otimes\mathcal{H}_\tau$ as the representation space of the product group $G\times G$ under the unitary representation $T:=\pi\otimes\tau$, defined as: 
\begin{equation}
T(g_1,g_2):=\pi(g_1)\otimes\tau(g_2).
\end{equation} 
Representation $T$ is irreducible as a representation of $G \times G$ \cite{reduc} and moreover, every irreducible representation of $G\times G$ is of that form, up to a unitary equivalence \cite{Folland}. Hence, we may view $G\times G$ as the kinematical group of the composite system. Since $G\times G$ is obviously compact, we can apply to it all the methods of Section II.

Let us consider a separable state $\varrho$ on $\mathcal{H}_\pi\otimes\mathcal{H}_\tau$, for which there exists a convex decomposition of the type (\ref{Nsep}): $\varrho = \sum_i p_i |u_i\rangle\langle u_i| \otimes |v_i\rangle\langle v_i|$. Then, from Eq. (\ref{phi}) we obtain that:
\begin{equation}\label{sep}
\phi_\varrho(g_1,g_2)=\sum_i p_i \kappa_i(g_1)\eta_i(g_2),
\end{equation}
where $\kappa_i(g_1):=\langle u_i |\pi(g_1) u_i\rangle$, $\eta_i(g_2):=\langle v_i |\tau(g_2) v_i\rangle$ are non-commutative characteristic functions from $\mathcal{P}_1(G)$, or more precisely from $\mathcal{E}_1(G)$. Conversely, a function of the form (\ref{sep}) defines a separable state through the integral (\ref{ro}), because:
{\setlength \arraycolsep{0pt}
\begin{eqnarray}
\int_{G\times G} \de g_1 \de g_2\,&\,d_T&\, \phi(g_1,g_2) T(g_1,g_2)^\dagger = \nonumber\\
&=&\sum_i p_i \Bigg( \int_G \de g_1 \,d_\pi \kappa_i(g_1) \pi(g_1)^\dagger\Bigg)\nonumber\\
& &\otimes\Bigg( \int_G \de g_2 \,d_\tau \eta_i(g_2) \tau(g_2)^\dagger\Bigg),\label{sep2}
\end{eqnarray}}
where $\de g: = \de g_1 \de g_2$ is the Haar measure on $G\times G$. Moreover, since we need to integrate in Eq. (\ref{sep2}) in order to obtain a density matrix, it is enough that the decomposition (\ref{sep}) holds almost everywhere w.r.t. the measure $\de g$. Hence we obtain the following theorem:

{\bf Theorem 1}. {\it Let $G$ be a compact kinematical group; $\pi$, $\tau$ its irreducible representations. A state $\varrho$ on $\h_\pi\otimes \h_\tau$ is separable if and only if its non-commutative characteristic function $\phi_\varrho$ can be written as a convex combination: $\phi_\varrho(g_1,g_2)=\sum_i p_i \kappa_i(g_1)\eta_i(g_2)$, where $\kappa_i, \eta_i \in \mathcal{E}_1(G)$ and the equality holds almost everywhere w.r.t. the Haar measure on $G\times G$.}

The above Theorem is our group-theoretical reformulation of the separability problem. The generalization to arbitrary multipartite systems is straightforward. We call the functions possessing decompositions of the type (\ref{sep}) {\it separable} and otherwise - {\it entangled}. One may thus generalize the separability problem to groups in the following way: 

{\bf Generalized separability problem}. {\it Given an arbitrary function $\phi\in\Po (G\times G)$, decide whether it is separable or not.}

This is an interesting mathematical problem, with connections to e.g. properties of polynomials on groups: if $G=SU(2)$, then, since $\phi_\varrho$ are polynomials in the group parameters, Theorem 1 states that a state is separable iff its group polynomial separates into two polynomials in the variables $g_1$ and $g_2$ respectively. 

One of the potential advantages of the current approach is its universality. For example, for $G=SU(2)$ characterization of separable functions within $\mathcal{P}_1(SU(2)\times SU(2))$ would lead through Eqs. (\ref{ro}) and (\ref{ro1}) to the characterization of all separable states in all possible finite dimensions. The other, more conceptual, advantage will be discussed in Section \ref{noncom}.
Note that if one considers a restriction $\phi|_{Abel}$ of an arbitrary $\phi\in\mathcal{P}_1(G\times G)$ to any Abelian subgroup of $G\times G$ (like Cartan subgroup if $G$ is a Lie group \cite{Zhelobienko}), then the separable decomposition (\ref{sep}), possibly infinite, always exists. This follows from the fact that on Abelian groups the usual Fourier transform is available. For a concrete example consider $G=SU(2)$. Then the maximal Abelian subgroup is $U(1)\times U(1)$ and one can always write: 
\begin{equation}
\phi(\theta_1,\theta_2)=\sum_{k,l} \hat{\phi}_{kl}\,\text{e}^{-\text{i}k\theta_1}\,\text{e}^{-\text{i}l\theta_2},\label{Fourier}
\end{equation}
where the angles $\theta_1,\theta_2$ parametrize $U(1)\times U(1)$ and $\hat{\phi}_{kl}$ are the Fourier coefficients of $\phi|_{U(1)\times U(1)}$. Since $\text{e}^{-\text{i}k\theta} \in \mathcal{P}_1(U(1))$, $\hat{\phi}_{kl}\ge 0$ by Bochner's Theorem \cite{Folland}, and $\sum_{kl}\hat{\phi}_{kl}=1$ by normalization of $\phi$, the Fourier series (\ref{Fourier}) is just the separable decomposition of $\phi|_{U(1)\times U(1)}$. For characteristic functions of states, i.e. for $\phi=\phi_\varrho$, the series (\ref{Fourier}) is finite: $k=-2j_1,-2j_1+2,\dots,2j_1$, $l=-2j_2,-2j_2+2,\dots,2j_2$, where $j_1$, $j_2$ are the corresponding spins, as $\phi_\varrho$'s are polynomials of bi-degree $(2j_1, 2j_2)$ in the group parameters; see Appendix A. However, for separable states the decomposition (\ref{Fourier}) will not generically prolong to the whole $SU(2)\times SU(2)$, as it contains at most $(2j_1+1)(2j_2+1)=mn$ terms, whereas from Caratheodory's Theorem we know that the number of terms in a separable decomposition is bounded by $m^2 n^2$ \cite{Pawel}. We further develop the connection between group non-commutativity and entanglement in Section \ref{noncom}.

\section{Analysis of the PPT criterion and pure states}
In Section II we have seen that the language of non-commutative characteristic functions is as valid description of quantum states as the usual language of density matrices. Hence, in particular, the known separability criteria should have their group-theoretical analogs. In this Section we show how to translate the PPT criterion \cite{PPT} and a simple criterion for pure states as two examples. Recall that the PPT condition implies that if $\varrho$ is separable, then the partially transposed matrix $\varrho^{T_1}$ is positive semidefinite \cite{chujka}.

Let us first note that for an arbitrary positive definite function $\phi$ it holds: 
\begin{equation}
\phi(g^{-1})=\overline{\phi(g)},\label{sraka}
\end{equation} 
and $\overline{\phi}$ is again positive definite. Hence, we immediately obtain from Eq. (\ref{sep}) a necessary separability criterion for an arbitrary $\phi\in \Po(G\times G)$: 
 
{\bf Proposition 1.} {\it If $\phi\in\mathcal{P}_1(G\times G)$ is separable then $\widetilde{\phi}(g_1,g_2):=\phi(g_1^{-1},g_2) \in\mathcal{P}_1(G\times G)$.}

In particular, from Theorem 1 we obtain the implication: ($\varrho$ - separable) $\Rightarrow \widetilde{\phi}_\varrho\in\mathcal{P}_1(G\times G)$. We will show that it is intimately related to the PPT condition. For that we will first consider $G=SU(2)$:

{\bf Proposition 2.} {\it $\widetilde{\phi}_\varrho\in\Po(SU(2)\times SU(2))$ if and only if $\varrho^{T_1}\ge 0.$} 

{\it Proof.} Let us first assume that $\varrho^{T_1}\ge 0$, so that $\phi_{\varrho^{T_1}}\in\mathcal{P}_1(SU(2)\times SU(2))$. The latter is just $\phi_{\varrho^{T_1}}(g_1,g_2)=\tr\big[\varrho \,\,\pi(g_1)^T\otimes \tau(g_2)\big]=\tr\big[\varrho \,\,\overline{\pi(g_1^{-1})}\otimes \tau(g_2)\big]$, and since for any $a\in SU(2)$
\begin{equation} \label{equiv}
\bar{a}=uau^{-1},\quad u:=-i\sigma_y \, , 
\end{equation}
and $\pi$ polynomially depends on the group parameters (see Appendix A), we obtain that $\phi_{\varrho^{T_1}}(g_1,g_2)=\phi_{\varrho}(ug_1^{-1}u^{-1},g_2)$. The condition (\ref{fdt}) for $\phi_{\varrho^{T_1}}$ takes then the following form:
{\setlength \arraycolsep{0pt}
\begin{eqnarray}
& & \iint \de \tilde{g} \,\de \tilde{h} \overline{f(\tilde{g})}\phi_{\varrho^{T_1}}(\tilde{g}^{-1}\tilde{h})f(\tilde{h})=\nonumber\\
& &=\int\! \de g_1 \, \de g_2\,\,\int\de h_1 \, \de h_2\,\,\overline{f(g_1 u,g_2)}\nonumber\\
& & \times \widetilde{\phi}_{\varrho}(g_1^{-1}h_1,g_2^{-1}h_2)f(h_1u,h_2)\ge 0\label{ppteqv},
\end{eqnarray}}
where $\tilde{g}:=(g_1,g_2)$. Since the inequality (\ref{ppteqv}) is satisfied for any $f\in L^1(SU(2)\times SU(2))$, the right shift by $u$ of the first argument is irrelevant. Thus, we get that  $\widetilde{\phi}_{\varrho}\in\mathcal{P}_1(SU(2)\times SU(2))$ (the normalization follows trivially).

On the other hand, let us assume that $\widetilde{\phi}_{\varrho}\in\mathcal{P}_1(SU(2)\times SU(2))$. Then from the similar argument to that leading to the condition (\ref{p3}), we can construct a positive semidefinite operator:
{\setlength \arraycolsep{0pt}
\begin{eqnarray}
\int_{G\times G} &\de g_1& \de g_2 \,\,d_T\, \widetilde{\phi}_\varrho(g_1,g_2) T(g_1,g_2)^\dagger=\nonumber\\ 
& &=\int\de g_1 \, \de g_2\,\,d_\pi d_\tau\, \phi_\varrho(g_1,g_2)\,\nonumber\\
& & \times\overline{\pi(u)} \big[\pi(g_1)^\dagger\big]^T\pi(u)^T\otimes \tau(g_2)^\dagger \nonumber\\
& &= \big[\overline{\pi(u)}\otimes {\bf 1}\big]\,\varrho^{T_1} \,\big[\pi(u)^T\otimes {\bf 1}\big] \ge 0, \label{pptlast}
\end{eqnarray}}
where in the first step we used the fact that $\de g^{-1}=\de g$. Since the local unitary rotation by $\pi(u)^T\otimes {\bf 1}$ does not affect the positivity of the operator in the inequality (\ref{pptlast}), the latter is equivalent to $\varrho^{T_1}\ge 0$. $\Box$

The crucial role in the above proof, especially in obtaining the inequality (\ref{pptlast}), has been played by the relation (\ref{equiv}), implying a unitary equivalence, denoted by $\sim\,$, between $SU(2)$-representations $\tau_k$ and their complex conjugates $\overline{\tau_k}$ for all $k$: $\overline{\tau_k}=C_k\tau_k C_k^\dagger$, i. e. $\tau_k \sim \overline{\tau_k}$. The intertwining isomorphisms $C_k$, equal to $\tau_k(u)$ for this particular group, satisfy $\overline{C_k}C_k={\bf 1}$. Representations with such properties are called representations of real type \cite{Trautman}. 

Now a natural question arises: if we consider a kinematical group which possesses at least one irreducible representation $\pi \not\sim\overline{\pi}$ (for example $G=SU(3)$, $\pi=\text{id}$), can we obtain from Proposition 1 any new criterion, independent from the PPT condition?  The negative answer provides the next Theorem:

{\bf Theorem 2.} {\it Let $G$ be a compact kinematical group; $\pi$, $\tau$ its irreducible representations. For any state $\varrho$ on $\mathcal{H}_\pi\otimes\mathcal{H}_\tau$, $\varrho^{T_1}\ge 0$ if and only if $\widetilde{\phi}_\varrho\in\mathcal{P}_1(G\times G)$.}

{\it Proof.} For a general group $G$ the property (\ref{equiv}) does not hold and we cannot use the previous technique. However, $\widetilde\phi_\varrho$ can be represented as follows:
\begin{equation}\label{fizda}
\widetilde\phi_\varrho(g_1,g_2)=\tr\big[\varrho^{T_1} \,\,\overline{\pi(g_1)}\otimes \tau(g_2)\big],
\end{equation}
so that $\widetilde\phi_\varrho$ becomes a non-commutative characteristic function of $\varrho^{T_1}$, treated as an operator acting on $\h_{\overline\pi}\otimes\h_\tau$. Since $\overline\pi$ is irreducible iff $\pi$ is, we can invert the transformation (\ref{fizda}):
\begin{equation}
\varrho^{T_1}=\int_{G\times G} \de g_1 \de g_2\,\,d_\pi d_\tau\, \widetilde{\phi}_\varrho(g_1,g_2)\, \overline\pi(g_1)^\dagger\otimes\tau(g_2)^\dagger.
\end{equation}
Then the statement follows immediately from the general results of Section II: if $\widetilde{\phi}_\varrho\in\mathcal{P}_1(G\times G)$, positivity of $\varrho^{T_1}$ follows from the same argument as that leading to the inequality (\ref{p3}). On the other hand, if $\varrho^{T_1}\ge 0$ then a direct calculation shows that $\widetilde\phi_\varrho$ satisfies the condition (\ref{fdt}). 
$\Box$

Let us now briefly examine pure states. For pure states a number of necessary and sufficient separability conditions is available. The one which is most easily translated into the group-theoretical language is the following: 

$\psi\in\mathcal{H}$ {\it is product} $\Leftrightarrow \tr_1(\tr_2 |\psi\rangle\langle \psi|)^2=1 = \tr_2(\tr_1 |\psi\rangle\langle \psi|)^2$. 

Using the orthogonality of matrix elements of representations, we easily obtain that this criterion is equivalent to the following integral condition:

{\bf Proposition 3}. {\it A function $\phi\in\mathcal{E}_1(G\times G)$ is product if and only if:
\begin{equation}\label{pure'}
\int_G\de g_1 \,\,d_\pi \, |\phi(g_1,e)|^2= 1 = \int_G\de g_2 \,\,d_\tau \, |\phi(e,g_2)|^2.
\end{equation}}

Note that the above condition applies to an {\it arbitrary} $\phi\in\mathcal{E}_1(G\times G)$ since, as we mentioned in Section II, every $\phi\in\mathcal{E}_1(G\times G)$ is of the form $\phi_\psi$ for some irreducible representation $\pi\otimes\tau$ of $G\times G$ and some pure state $\psi\in\h_\pi\otimes\h_\tau$. 

\section{Analysis on finite groups}
\label{finite}
In this Section we study the special case of finite kinematical groups. An example of such groups are symmetric groups $\mathfrak{S}_M$ (group of permutations of $M$ elements) and moreover, every finite group is isomorphic to a subgroup of some $\mathfrak{S}_M$ \cite{Zhelobienko}. Finite groups are in particular compact, and hence all the previous theory applies to them as well, with the only change being:
\begin{equation}
\int_G \de g \to \frac{1}{|G|}\sum_{g\in G},
\end{equation}
where $|G|$ is the number of elements of $G$ (its order). However, for finite groups several simplifications occur. First of all, if $|G|=N$, then the space of complex functions on $G$ is isomorphic to $\mathbb{C}^N$, and we may identify each function $\phi$ with a row vector $\vec{\phi}$ of its values. The positive definiteness condition (\ref{fdt}) takes then the following form:
\begin{equation}
\sum_{\alpha,\beta=1}^N \overline{c_\alpha}\phi(g_\alpha^{-1}g_\beta)c_\beta\ge 0 \quad \text {for any}\ \ \vec{c}\in\mathbb{C}^N, \label{pd}
\end{equation}
(indices $\alpha,\beta,\dots$ now enumerate the group elements), which is just the positive semidefiniteness condition for the matrix: 
\begin{equation}
{\bm \Phi}_{\alpha\beta}:=\phi(g_\alpha^{-1}g_\beta)\label{cipa}
\end{equation}
(compare with \cite{l1}). To closer examine the structure of this matrix, let us first fix the labelling of the group elements such that $g_1:=e$. Then, the first row of ${\bm \Phi}$ contains the values of the function $\phi$ itself, and hence, it determines the rest of the matrix. We may define a function $\sigma$ on $\mathbb{N}\times \mathbb{N}$ through: 
\begin{equation}
g_{\sigma(\alpha , \beta)}:=g_\alpha^{-1}g_\beta. \label{sigma}
\end{equation}
Note that $\sigma$ is completely determined by the group multiplication table, $\sigma(\alpha , \alpha)=1$, and it satisfies the cocycle condition:
\begin{equation}
g_{\sigma(\alpha , \beta)}g_{\sigma(\beta , \gamma)}=g_{\sigma(\alpha , \gamma)}\end{equation}
(no summation over $\beta$ here). Combining Eq. (\ref{sigma}) with the normalization condition (\ref{norm}), and the property (\ref{sraka}), we obtain a general form of the matrix (\ref{cipa}) for an arbitrary $\phi\in \Po(G)$:
\begin{equation}
{\bm \Phi} = \left[ 
\begin{array}{ccccc} 1 & \phi_2 & \phi_3             & \cdots & \phi_N \\ 
  \overline{\phi_2} &       1   & \phi_{\sigma(2,3)} & \cdots & \phi_{\sigma(2,N)}\\  
  \overline{\phi_3} & \overline{\phi_{\sigma(2,3)}} & 1 & \cdots & \phi_{\sigma(3,N)}\\
  \vdots               &   \vdots  & \vdots                & \ddots & \vdots\\
  \overline{\phi_N} & \overline{\phi_{\sigma(2,N)}}  & \overline{\phi_{\sigma(3,N)}}  & \cdots & 1 \end{array}
\right].\label{ochujalosc}
\end{equation}
In other words, the matrix ${\bm \Phi}$ is built from the vector $\vec{\phi}$ by permuting in each row (or column) its components according to the multiplication table of $G$. Relabelling of the group elements corresponds to unitary rotation of ${\bm \Phi}$, which does not affect the condition (\ref{pd}), and hence we may work with a fixed labelling. 

For pure states, one can rewrite the condition (\ref{czysta}) in the following form:\begin{equation} 
\phi_\varrho(g^{-1}g')=\int_G \de h \,d_\tau \,\phi_\varrho(g^{-1}h)\phi_\varrho(h^{-1}g'),
\end{equation} 
from which it follows that $\varrho$ is pure iff 
\begin{equation}
{\bm \Phi}^2_\varrho=\frac{N}{d_\tau}{\bm \Phi}_\varrho.
\end{equation}
Hence, $(d_\tau/N) {\bm \Phi}_\varrho$ is a projector. 

Let us now move to bipartite systems, i. e. to systems with the kinematical group $G\times G$. We may view functions $\phi$ on such group either as $N\times N$ matrices $\phi_{\alpha\beta}:=\phi(g_\alpha , g_\beta)$, or as vectors from $\mathbb{C}^{N^2}$. The separability criterion --- Theorem 1 --- takes then the following form on finite $G$:

{\bf Proposition 4.} {\it A function $\phi\in\Po (G\times G)$ is separable if and only if there exists a convex decomposition:
\begin{equation}
\phi_{\alpha\beta}=\sum_i p_i\, \kappa_{i\alpha} \,\eta_{i\beta},\label{zdzira}
\end{equation}
where for each $i$ vectors $\vec{\kappa_i}, \vec{\eta_i}\in\mathbb{C}^{N}$ lead, according to the prescription (\ref{ochujalosc}), to positive semidefinite matrices.}

Decomposition (\ref{zdzira}) resembles the singular value decomposition of the matrix $\phi_{\alpha\beta}$, however the vectors are specifically constrained. Let us mention another, equivalent form of Proposition 4:

{\bf Proposition 5.} {\it A function $\phi\in\Po (G\times G)$ is separable if and only if its matrix ${\bm \Phi}$ defined by Eq. (\ref{cipa}) can be convexly decomposed as follows:
\begin{equation}
{\bm \Phi}=\sum_i p_i {\bm K}_{i} \otimes {\bm N}_{i},\label{zdzira2}
\end{equation}
where for each $i$, ${\bm K}_{i}, {\bm N}_{i} \ge 0$ and are of the form (\ref{ochujalosc}) for some $\vec{\kappa_i}, \vec{\eta_i}\in\mathbb{C}^{N}$.}

The proof follows for the fact that the first row of the matrix equality (\ref{zdzira2}) is just the Eq. (\ref{zdzira}), and from the specific structure (\ref{ochujalosc}) of the matrices in Eq. (\ref{zdzira2}). 

>From the condition (\ref{pd}), the matrix ${\bm \Phi}_{\alpha\alpha ', \beta\beta '}=\phi(g_\alpha^{-1}g_\beta, g_{\alpha '}^{-1}g_{\beta '})$ is positive semidefinite as an operator on $\mathbb{C}^{N}\otimes\mathbb{C}^{N}$, and, after rescaling by $1/N^2$, has trace one. Hence, Proposition 5 embeds the given separability problem into the higher dimensional one \cite{alechuj}. Note however that the matrices in Eq. (\ref{zdzira2}) are of a very specific form: they are completely determined by their first rows and the group multiplication table. The necessary separability criterion from Proposition 1 takes a particularly familiar from for finite groups:

{\bf Proposition 6.} {\it If $\phi\in\mathcal{P}_1(G\times G)$ is separable then ${\bm \Phi}^{T_1}\ge 0$.}

The proof follows from Proposition 1, the equality: $\widetilde{{\bm \Phi}}_{\alpha\alpha ', \beta\beta '}=\widetilde{\phi}(g_\alpha^{-1}g_\beta, g_{\alpha '}^{-1}g_{\beta '})=\phi(g_\beta^{-1}g_\alpha, g_{\alpha '}^{-1}g_{\beta '}) ={\bm \Phi}_{\beta\alpha ', \alpha\beta '} $, and the positive definiteness condition (\ref{pd}). 

\section{Formal resemblance to local hidden variable models}
\label{lhv}
Let us here remark on a purely formal resemblance of the group-theoretical formalism from the preceding sections to LHV models \cite{WernerWolf}. Following the usual approach, let us consider an expectation value of a product operator $A\otimes B$, where $A=\sum_\mu a_\mu P_\mu$, $B=\sum_\nu b_\nu Q_\nu$ are the corresponding spectral decompositions. Using the representation (\ref{ro}), the mean value of $A\otimes B$ in the state $\varrho$ can be written as follows:
{\setlength \arraycolsep{0pt}
\begin{eqnarray}
\tr(A\otimes B\varrho) &=& \sum_{\mu,\nu} a_\mu b_\nu \int_{G\times G}\de g_1 \, \de g_2\,\,d_\pi d_\tau\, \phi_\varrho(g_1,g_2)\nonumber\\
& &\times \tr\big[P_\mu\pi(g_1)^\dagger\big]\,\tr\big[Q_\nu \tau(g_2)^\dagger\big]. 
\end{eqnarray}}
Hence, the probability $p(\mu,\nu|A,B)$ of obtaining the value $a_\mu$ for $A$ and $b_\nu$ for $B$ is given by:
{\setlength \arraycolsep{0pt}
\begin{eqnarray}
p(\mu,\nu|A,B)&=&\int\de g_1 \, \de g_2\,\,d_\pi d_\tau\, \phi_\varrho(g_1,g_2)\nonumber\\
& &\times \tr\big[P_\mu\pi(g_1)^\dagger\big]\,\tr\big[Q_\nu \tau(g_2)^\dagger\big].
\end{eqnarray}}
This expression {\it formally} resembles a LHV model, where the role of the probability space plays $G\times G$, the ``response functions'' are $R(\mu,g_1):=\tr\big[P_\mu\,\pi(g)^\dagger\big]$ and $R(\nu,g_2):=\tr\big[Q_\nu\, \tau(g_2)^\dagger\big]$, and the ``probability measure'' is $\de m := d_\pi d_\tau \,\phi_\varrho(g_1,g_2)\,\de g_1 \de g_2$. The resemblance is of course only formal, since the ``response functions'', as well as the measure $\de m$, are complex. The response functions satisfy only $R(g^{-1})=\overline{R(g)}$, while the measure $\de m$ is positive definite but not positive.

\section{Non-commutativity and entanglement}
\label{noncom}
We conclude with a general remark, connecting the existence of entanglement with non-commutativity of the kinematical group $G$. For that we first have to change the usual mathematical language of quantum statistics (we do not consider dynamics here). Instead of using Hilbert spaces and density matrices, let us: i) assume that the kinematical arena is set up by the kinematical group $G$; ii) represent physical states by functions from $\mathcal{P}_1(G)$ (or its subset), rather than by density matrices; iii) for composite systems, take as the kinematical group the product group $G\times G \times G \dots$ (for alternative group-theoretical reformulations see e.g. Refs. \cite{Mielnik, Kus, Naudts}). As we have seen in Section II, such a description is indeed equivalent to the standard one, provided that the kinematical group is chosen correctly: for spin systems $G=SU(2)$, for canonically quantized particles it is the Heisenberg-Weyl group \cite{chujnia}, while for classical particles $G$ is just the phase-space $\mathbb{R}^{2n}$.  

Now, let us assume that the kinematical group is Abelian. Then by Bochner's Theorem \cite{Folland}, our states, i.e. functions from $\mathcal{P}_1(G)$, are in one-to-one correspondence with Borel probability measures on the space of all irreducible representations of $G$, $\hat{G}$, which in this case is also an Abelian group (for instance $\widehat{\mathbb{R}^{2n}} \simeq \mathbb{R}^{2n}$). Hence, we recover classical statistical description of our system \cite{Mackey}, with  $\hat{G}$ playing the role of the phase-space (at least for the purpose of statistics). If, moreover, the system under consideration is multipartite, then due to the fact that $\widehat{G_1 \times G_2}=\hat{G_1}\times\hat{G_2}$ \cite{Folland}, the phase-space of the composite system is the usual Cartesian product of the individual phase-spaces, and our states correspond to the probability measures on this product. There is no place for entanglement here, understood as the impossibility of generating the composite system state-space from the individual state-spaces, because probability measures on Cartesian products can always be decomposed (under suitable limits) into the convex mixtures of product measures (due to the underlying structure of the $\sigma$-algebra of Borel sets).

On the other hand, when $G$ is non-Abelian, then Bochner's Theorem cannot be applied, and $\mathcal{P}_1(G)$ is in one-to-one correspondence with density matrices through the inverse Frourier transform (\ref{ro}) and (\ref{ro1}). Since density matrices exhibit entanglement, one may view the latter as the consequence of the non-commutativity of the kinematical group $G$. The last observation opens some possibility of speculations on the connection between entanglement and the uncertainty principles. In this context we note that G\"uhne has developed in Ref. \cite{Otfredo} some methods of entanglement description with the help of uncertainty relations.

Let us also mention that the general group-theoretical approach, sketched above, can be also applied to canonically quantized systems and the analysis of the correspondence principle \cite{chujnia}. 

Finally, as discussed at the end of Appendix A, our approach opens a possibility of deriving highly non-trivial statements on positive definite functions on product groups, using theory of entanglement.

We would like to thank E. Bagan, J. Kijowski, J. Wehr, K. $\dot{\text{Z}}$yczkowski and especially M. Ku\'s  for discussions, and the  Deutsche Forschungsgemeinschaft
(SFB 407, 436 POL), ESF PESC QUDEDIS, EU IP Programme ``SCALA'', and MEC (Spanish Goverment) under contract FIS2005-04627 for the financial support. 

\appendix
\section{$SU(2)$-characteristic function of the $3\otimes 3$ Horodecki's state}
As an example we calculate for $G=SU(2)$ the characteristic function of the $3\otimes 3$ PPT entangled state, discovered by P. Horodecki in Ref. \cite{Pawel}. Since any irreducible representation $T$ of $SU(2)\times SU(2)$ is of the form $T=\tau_{j_1}\otimes \tau_{j_2}$ for some spins $j_1$, $j_2$, all we need are the matrix elements $\tau^j_{\mu\nu}$ of the corresponding spin-$j$ representations $\tau_j$ of $SU(2)$. The concrete basis $\{e_\mu\}$ in which we calculate them is irrelevant for our purposes, as from Eq. (\ref{phi}) it follows that a change of basis: $\tau_{j_1} \mapsto U_1\tau_{j_1}U_1^\dagger$, $\tau_{j_2} \mapsto U_2\tau_{j_2}U_2^\dagger$ induces only a local rotation of the state $\varrho$:
\begin{equation}
\tr\big[\varrho U_1\tau_{j_1}U_1^\dagger\otimes U_2\tau_{j_2}U_2^\dagger\big]=\tr\big[\big(U_1^\dagger\otimes U_2^\dagger\varrho U_1\otimes U_2\big)\tau_{j_1}\otimes\tau_{j_2}\big],
\end{equation}
and the rotated state $U_1^\dagger\otimes U_2^\dagger\varrho U_1\otimes U_2$ is separable iff $\varrho$ is separable. The above remark concerning bases obviously  applies to any kinematical group $G$.

A convenient formula for $\tau^j_{\mu\nu}$ can be found, for example, in Ref. \cite{Zhelobienko}:
\begin{equation}
\tau^j_{\mu\nu}(g)=\frac{1}{(j-\mu)!}\frac{\de ^{j-\mu}}{\de z^{j-\mu}}\bigg|_0\big[(\alpha z +\beta)^{j-\nu}(-\overline{\beta}z+\overline{\alpha})^{j+\nu}\big],\label{elementa}
\end{equation}
where $\mu,\nu = -j,-j+1,\dots , j$ and $\alpha$, $\beta$ are the group parameters:
\begin{equation}
g=\left[   \begin{array}{cc} \alpha & -\overline{\beta}\\
                       \beta  &  \overline{\alpha} \end{array} \right],\quad |\alpha|^2+|\beta|^2=1 .
\end{equation}
>From Eq. (\ref{elementa}) we immediately see that matrix elements of the representation $\tau_j$ are homogeneous polynomials of degree $2j$ in the group parameters. Hence, matrix elements of $\tau_{j_1}\otimes\tau_{j_2}$ are polynomials of bi-degree $(2j_1, 2j_2)$ in $(\alpha_1, \beta_1)$, $(\alpha_2,\beta_2)$ as mentioned in Section II. 

The $3 \otimes 3$ Horodecki's state is given by:
\begin{eqnarray}\label{3x3}
\varrho = \frac{1}{8a+1}\!\!
         \left[   \begin{array}{ccccccccc} a & 0 & 0 & 0 & a & 0 & 0 & 0 & a\\
                                           0 & a & 0 & 0 & 0 & 0 & 0 & 0 & 0\\
                                           0 & 0 & a & 0 & 0 & 0 & 0 & 0 & 0\\
                                           0 & 0 & 0 & a & 0 & 0 & 0 & 0 & 0\\
                                           a & 0 & 0 & 0 & a & 0 & 0 & 0 & a\\
                                           0 & 0 & 0 & 0 & 0 & a & 0 & 0 & 0\\
                                           0 & 0 & 0 & 0 & 0 & 0 & \frac{1+a}{2} & 0 & \frac{\sqrt{1-a^2}}{2}\\
                                           0 & 0 & 0 & 0 & 0 & 0 & 0 & a & 0\\
                                           a & 0 & 0 & 0 & a & 0 & \frac{\sqrt{1-a^2}}{2} & 0 & \frac{1+a}{2}\\ 
                       \end{array}\right]\!\!,
\end{eqnarray}
where $0\le a \le 1$. From Eq. (\ref{elementa}) we find the three dimensional representation of $SU(2)$:
\begin{equation}\label{tau_1}
\tau_1(g)=\left[   \begin{array}{ccc} \alpha^2 & -\alpha\overline{\beta} & \overline{\beta}^2\\
                                      2\alpha\beta & |\alpha|^2-|\beta|^2 & -2\overline{\alpha\beta}\\
                                      \beta^2 & \overline{\alpha}\beta & \overline{\alpha}^2 \end{array} \right].
\end{equation}
Inserting Eqs. (\ref{3x3}) and (\ref{tau_1}) into Eq. (\ref{phi}), we obtain the characteristic function of the state (\ref{3x3}):
{\setlength \arraycolsep{0pt}
\begin{eqnarray}
\phi_\varrho(g_1,g_2)&=&\frac{a}{8a+1}\bigg[(\alpha_1^2+\frac{1}{2}\overline{\alpha_1}^2)(\alpha_2^2+\overline{\alpha_2}^2)+(\beta_1\beta_2)^2\nonumber\\
&+& (\overline{\beta_1}\overline{\beta_2})^2+4\alpha_1\beta_1\alpha_2\beta_2+4\overline{\alpha_1\beta_1\alpha_2\beta_2}\nonumber\\
&+& \alpha_1\overline{\beta_1}\alpha_2\overline{\beta_2} + \overline{\alpha_1}\beta_1\overline{\alpha_2}\beta_2\nonumber\\ 
&+& (\alpha_1^2+\overline{\alpha_1}^2)(|\alpha_2|^2-|\beta_2|^2)\nonumber\\
&+& (|\alpha_1|^2-|\beta_1|^2)(\alpha_2^2+\overline{\alpha_2}^2)\label{pierd}\\
&+& (|\alpha_1|^2-|\beta_1|^2)(|\alpha_2|^2-|\beta_2|^2)\bigg]\nonumber\\
&+& \frac{\sqrt{1-a^2}}{2}\overline{\alpha_1}^2(\beta_2^2+\overline{\beta_2}^2)
+ \frac{1}{2}\overline{\alpha_1}^2(\alpha_2^2+\overline{\alpha_2}^2)\nonumber.
\end{eqnarray}}
Note that from the fact that the state (\ref{3x3}) is entangled for $0 < a < 1$, we obtain through Theorem 1, a highly non-trivial result concerning the function (\ref{pierd}): the function (\ref{pierd}) cannot be represented as a convex mixture of products of positive definite functions, depending on parameters $(\alpha_1, \beta_1)$, and  $(\alpha_2,\beta_2)$ respectively.


\begin{thebibliography}{99}
\bibitem{l1} Since $\phi$ is continuous, the condition (\ref{fdt}) is equivalent to the following one: for any $N$ and any sequence $\{g_\alpha \}\in G$, $\alpha=1,\dots,N$ matrix $\phi(g_\alpha^{-1}g_\beta)$ is positive semidefinite; see Ref. \cite{Folland} for more details.
\bibitem{ort} Peter-Weyl Theorem states, among the others, that: i) for any irreducible representation $\tau$ of a compact group $G$, holds: $\int \de g d_\tau \tau_{\mu\nu}(g) \overline{\tau_{\mu' \nu'}(g)} = \delta_{\mu\mu'}\delta_{\nu\nu'}$, where matrix elements $\tau_{\mu\nu}:=\langle e_\mu |\tau e_\nu\rangle$ of $\tau$ are taken w.r.t. any orthonormal basis $\{e_\mu \}$ of $\h_\tau$; ii) for any two non-equivalent irreducible representations $\pi, \tau$ of $G$ their matrix elements are orthogonal: $\int \de g \pi_{\mu\nu}(g) \overline{\tau_{\mu' \nu'}(g)} = 0$; see e.g. Refs. \cite{Folland, Zhelobienko}.
\bibitem{GNS} One can also recover density matrices from their non-commutative characteristic functions using the GNS construction \cite{Folland}, which states that to each $\phi\in\mathcal{P}_1(G)$ one can assign a Hilbert space $\mathcal{H}_\phi$, a (generically reducible) representation $\tau_\phi:G\to U(\mathcal{H}_\phi)$, and a cyclic vector $v_\phi$, such that: $\phi(g)=\langle v_\phi |\tau_\phi(g) v_\phi \rangle$. If $\phi=\phi_\varrho$, so that $\phi$ is of the form (\ref{span}), then $v_\phi$ is the, so called, purification of $\varrho$. Although deeper and more general, the GNS construction is less explicit than the inverse non-commutative Fourier transform (\ref{ro}).
\bibitem{reduc} Obviously $T$ becomes reducible when restricted to the diagonal subgroup $\Delta:=\{(g,g);g\in G\}$, but this does not concern us here.
\bibitem{chujka} For definiteness' sake, we transpose only w.r.t. the first party but our arguments, after obvious changes, hold equally well for transposition w.r.t. the second party too.
\bibitem{alechuj} For example, if one wants to study $3 \otimes 3$-dimensional separability problem, then the lowest order finite group, which possesses a three dimensional irreducible representation is the alternating group $A_4$ (the group of even permutations from $\mathfrak{S}_4$). Its order is $N=12$ and Eq. (\ref{zdzira2}) states then the specifically symmetric separability problem in the dimension $12\otimes 12$.
\end{thebibliography}
\end{document}